\documentclass[aps,prd,preprint,tightenlines,groupedaddress,nofootinbib,showpacs,byrevtex]{revtex4}
\usepackage{amssymb,latexsym}
\usepackage{amsmath,amsbsy}
\usepackage{epsfig,bm}
\usepackage{graphicx,comment}
\DeclareMathOperator{\tr}{tr}

\begin{document}
\def\a{{\alpha}}
\def\b{{\beta}}
\def\d{{\delta}}
\def\D{{\Delta}}
\def\e{{\varepsilon}}
\def\g{{\gamma}}
\def\G{{\Gamma}}
\def\k{{\kappa}}
\def\l{{\lambda}}
\def\L{{\Lambda}}
\def\m{{\mu}}
\def\n{{\nu}}
\def\o{{\omega}}
\def\O{{\Omega}}
\def\S{{\Sigma}}
\def\s{{\sigma}}
\def\th{{\theta}}

\def\ol#1{{\overline{#1}}}

\def\Dslash{D\hskip-0.65em /}

\def\CPT{{$\chi$PT}}
\def\QCPT{{Q$\chi$PT}}
\def\PQCPT{{PQ$\chi$PT}}
\def\tr{\text{tr}}
\def\str{\text{str}}
\def\diag{\text{diag}}
\def\order{{\mathcal O}}

\def\cC{{\mathcal C}}
\def\cB{{\mathcal B}}
\def\cT{{\mathcal T}}
\def\cQ{{\mathcal Q}}
\def\cL{{\mathcal L}}
\def\cO{{\mathcal O}}
\def\cA{{\mathcal A}}
\def\cQ{{\mathcal Q}}
\def\cR{{\mathcal R}}
\def\cH{{\mathcal H}}
\def\cW{{\mathcal W}}
\def\cM{{\mathcal M}}
\def\cJ{{\mathcal J}}
\def\cF{{\mathcal F}}
\def\cK{{\mathcal K}}
\def\cY{{\mathcal Y}}

\def\eqref#1{{(\ref{#1})}}
 
\title{Mixed Action Lattice Spacing Effects on the Nucleon Axial Charge}
\author{ F.-J.~Jiang}
\email[]{fjjiang@itp.unibe.ch}
\affiliation{Institude of Theoretical Physics\\
Sidlerstrasse 5\\
Ch-3012 Bern\\
Switzerland}

\date{\today}


\begin{abstract}
We study the nucleon axial charge in the chiral perturbation theory for a 
mixed lattice action of Ginsparg-Wilson valence quarks and staggered sea 
quarks. In particular, we investigate the lattice spacing $a^2$-dependence 
of the neutron to proton axial transition matrix element. By using the
known low-energy constants (LEC's) and an estimated value of a new LEC which 
appears in the calculation, we see a large lattice spacing effect 
on $g_{A}$.   
  
\end{abstract}

\maketitle

\section{Introduction}

Lattice QCD, the only known first principle non-perturbative approach to QCD, 
has made a dramatic progress in the last decade because of the increasing 
power in the computing resources and the improvement of the 
numerical algorithms for the simulations. Today fully dynamical lattice QCD 
simulations together with effective theories enable one to make first 
principle predictions of hadronic physics. Recently, a so-called 
mixed-action lattice simulation 
using different kind of fermions in the valence and sea sectors has gained
attention because these simulations can reach the pion masses as low
as $300{\text{MeV}}$ \cite{Aubin:2004fs,Aubin:2004ej,Silas:2005,Edward:2005,
Silas:2006.1,Silas:2006.2,Silas:2006.3,Silas:2006.4,Edward:2005.1,Negele:2006.1,Negele:2006.2}. In particular, despite 
the controversy of the validity of the fourth root trick used to 
overcome the fermion taste doubling problem of the staggered fermions,
the lattice calculations using Ginsparg-Wilson valence quarks in 
staggered sea quarks has become popular
because many quantitative results are obtained 
\cite{Aubin:2004fs,Aubin:2004ej,Silas:2005,Edward:2005,
Silas:2006.1,Silas:2006.2,Silas:2006.3,Silas:2006.4}. Further, 
the publicly available MILC configurations 
\cite{Bernard:2001av} have encouraged the
use of lattice calculations with dynamical staggered fermions.
   
The lattice action can be described by Symanzik 
action~\cite{Symanzik:1983dc,Symanzik:1983gh} which is based on the 
symmetries of the underlying lattice theory and can be
organized in powers of the lattice spacing $a$:
\begin{equation}
\cL_{\text{Sym}} = 
\cL 
+ 
a \, \cL^{(5)} 
+ 
a^2 \, \cL^{(6)} 
+
\dots \label{eq:syman}
\, ,\end{equation}
where $\cL^{(n)}$ represents the contribution from dimension-$n$ 
operators. The symmetries of the lattice action are respected 
by the Symanzik Lagrangian $\cL_{\text{Sym}}$ order-by-order in $a$.
In order to address lattice spacing corrections, based on the Symanzik 
action, several mixed-action chiral perturbation
theories are developed to investigate the systematic errors which 
arise from lattice simulations due to the non-vanishing lattice spacing  
\cite{Bar:2002nr,Bar:2003mh,Silas:2003,bct0501,OBar:2005,bct0508}. These mixed-action chiral
perturbation theories provide a way to test the results from the lattice 
calculations using the mixed lattice action and vice-versa.    

The nucleon axial charge, $g_A$, is an important quantity in QCD which 
quantifies the spontaneous chiral symmetry breaking. Since its value
is known to a very high precision in experiments \cite{PDG:2006}, it can be 
used as a very good test for the first principle lattice calculations. 
This quantity of hadronic physics has been studied extensively 
in both lattice simulations 
and chiral perturbation theory \cite{Edward:2005,Jenkins:1991jv,Jenkins:1991es,
Blum:2004,Khan:2005,Borasoy:1999,Zhu:2001,s&m:2002,s&m:2004}. Although to perform the lattice
simulations to calculate $g_A$ is straightforward, some controversy remains. 
For example,
the $g_A$ obtained from lattice simulations has raised the controversy of  
the possibility of large volume effects in the chiral limit
\cite{Jaffe:2002,Cohen:2002}.
Recently, using the mixed-action of Domain Wall valence quarks with 
staggered sea quarks, \cite{Edward:2005} obtained a nice result of $g_A$
which seems to match nicely  
with the prediction from chiral perturbation theory 
and it is expected that the future data might be able to make contact 
with experiment via extrapolation. On the other hand, 
in \cite{Edward:2005}, the discretization errors 
in computing $g_A$ using the mixed-action might be overlooked. 
With the advent of mixed-action $\chi$PT \cite{OBar:2005} and extension
to the baryon sector \cite{bct0508}, it is surprising that
lattice spacing effects on the nucleon axial charge have
not been addressed. In this paper we address this issue mentioned above, 
namely, we use the mixed-action chiral perturbation theory to study 
the lattice-spacing dependence of $g_A$. In particular,
we found that it depends on the unphysical low energy constant $g_{1}$
and a new unknown LEC for which the actual value must be obtained from lattice
calculations. 

This paper is organized as follow. In Section
$2$, we briefly review the chiral perturbation theory of 
Ginsparg-Wilson valence quarks in staggered sea. In particular, we will put 
emphasis on the lattice-spacing $a$ dependence of the mixed-action 
Lagrangian. Next, in Section $3$, we will construct the axial-vector current 
from the mixed-action Lagrangian and compute the $pn$ matrix element of the 
axial-vector current. Further we will compare the results obtained in the 
presence and absence of lattice-spacing $a$. Finally we conclude 
in Section $4$. For completeness, we include some standard functions 
which arise in this calculation in the Appendix.

\section{Chiral Lagrangian}

To address the finite lattice spacing issue, with the pioneering 
work in \cite{Sharpe:1998xm,Lee:1999zx}, 
\cite{Rupak:2002sm,Bar:2002nr,Bar:2003mh} have extended 
the $\chi$PT in the meson sector. This
finite lattice spacing artifacts has also be investigated in staggered
$\chi$PT for the mesons 
\cite{Aubin:2003mg,Aubin:2003uc,Sharpe:2004is} and 
the mixed-action PQ$\chi$PT \cite{Silas:2003,OBar:2005,bct0501,bct0508}. 
To address the finite lattice spacing effects, one 
utilizes a dual expansion in the quark masses and lattice spacing
with the usual energy scale \cite{Silas:2003,Bar:2003mh}:
\begin{equation}
m_{q} \ll \Lambda_{QCD} \ll \frac{1}{a}\, ,
\end{equation} 
and the following power counting scheme \cite{bct0508}:
\begin{equation} \label{eqn:pc}
\e^2 \sim 
\begin{cases}
 m_q / \L_{QCD} \\
 a^2   \L_{QCD}^2
\end{cases}
\, ,\end{equation}
which is relevant for current improved staggered quarks simulations. 

In this section, we briefly introduce the chiral perturbation theory of 
the partially-quenched mixed-action theory. In particular, we focus on the
mixed-action of Ginsparg-Wilson valence quarks in staggered sea quarks. 
We only write the chiral Lagrangian of the associated mixed-action 
theory and do not go into the detail of how this chiral Lagrangian can be 
constructed from the Symanzik Lagrangian since the detail of the procedure
can be found in the references cited above. 
For our purpose in this paper, we only 
need to keep in mind that $\cL^{(5)} = 0$ and the taste-symmetry breaking and 
$SO(4)$ breaking operators will not enter at the order we calculate 
the axial-vector current matrix element \cite{bct0508}.

\subsection{Mesons}
In the following, the strange quark mass is assumed to be fixed to its 
physical value, therefore one can use two-flavor theory and does not
need to worry about the extrapolation in the strange quark mass.
The lattice action we consider here is built from $2$ flavors of 
Ginsparg-Wilson 
valence quarks and $2$ flavors of staggered sea quarks. In the continuum 
limit, the Lagrangian is just the partially quenched Lagrangian which is 
given by:
\begin{equation}
\cL = \ol Q \Dslash \, Q + \ol Q m_q Q\, ,
\end{equation} 
where the quark fields appears as a vector $Q$ with entries given by:
\begin{equation}
Q = ( u, d, j_1, j_2, j_3, j_4, l_1, l_2, l_3, l_4, \tilde{u}, \tilde{d},
 )^T\,
.\end{equation}
and transforms in the fundamental representation of the graded group
$SU(10|2)$.
Notice the fermion doubling has produced four tastes for each flavor $(j,l)$
of staggered sea quark. The partially quenched generalization of the mass 
matrix $m_q$ in the isospin limit is given by:
\begin{equation}
m_q = \diag (m_u, m_u, m_j \xi_I, m_j \xi_I, m_u, m_u) 
\, ,\end{equation}
with $\xi_I$ as the $4$ x $4$ taste identity matrix. 
The low-energy effective theory of the theory we consider above is written in 
terms of the pseudo-Goldstone mesons emerging from spontaneous symmetry 
breaking which are realized non-linearly in an $U(10|2)$ matrix exponential
$\Sigma$
\begin{equation}
 \Sigma=\exp\left(\frac{2i\Phi}{f}\right)
  = \xi^2\,.
\label{sig}
\end{equation} 
To the order (next-to-leading order) we work in investigating 
the 
$a^2$-dependence of the matrix element of the axial-vector current between 
nucleons, 
the relevant partially-quenched chiral perturbation theory Lagrangian for the 
mesons up to order $O(\epsilon^2)$ is given by:

\begin{equation} \label{eq:Llead}
\cL =  
\frac{f^2}{8}
\str \left( \partial_\mu\Sigma^\dagger \partial_\mu\Sigma\right)
    - \l \,\str\left(m_q\Sigma^\dagger+m_q^\dagger\Sigma\right)
    + \frac{1}{6} \mu_0^2 \, (\str \, \Phi)^2
    + a^2 \mathcal{V}\, .
\end{equation}
where
\begin{equation}
  \Phi=
    \left(
      \begin{array}{cc}
        M & \chi^{\dagger} \\ 
        \chi & \tilde{M}
      \end{array}
    \right)
\, ,\end{equation}
$f=132$~MeV, the str() denotes a graded flavor trace and
$\Sigma$ is defined in (\ref{sig}).  The $M$, 
$\tilde{M}$, and $\chi$ are matrices of pseudo-Goldstone bosons and 
pseudo-Goldstone fermions, for example, see \cite{s&m:2002}. 
The potential $\mathcal{V}$ contains the effects of dimension-$6$ operators
in the Symanzik action \cite{OBar:2005}. Expanding the Lagrangian in 
Eq.~\eqref{eq:Llead} to the leading order, one can determine the meson masses 
needed for the calculations of baryon observables. In particular, the 
relevant mesons needed for the axial-vector current matrix element 
calcualtions are the valence pion, mesons made 
of $S_i \overline{V}$ with one staggered sea quark $S_{i}$ of flavor $S$ and 
quark taste $i$ and a Ginsparg-Wilson valence quark $\overline{V}$ and 
finally mesons 
with two staggered quarks in a flavor-neutral, taste-singlet combination. 
The associated masses to the lowest order for the latter two mesons can 
be written in terms of the valence pion mass $m^2_{VV}$ which can be 
determined from the valence spectroscopy and the pseudoscalar taste 
pion mass $m^2_{SS}$ which the mass can be learned from the MILC 
spectroscopy and are given by:
\begin{eqnarray}
&m_{S V}^2& = \frac{m^2_{SS}+m^2_{VV}}{2} + 
\frac{16 \, a^2 \, C_{\text{mix}}}{f^2}\, ,\nonumber \\
&m_{SS,I}^2& = m^2_{SS}  + \frac{64 \, a^2}{f^2} (C_3 + C_4)\,
\label{qmass}
 ,\end{eqnarray}
 where $C_{\text{mix}}$ and $C_3+C_4$ are the parameters in the potential  
$\mathcal{V}$. Since these masses are independent of the quark taste, we do 
not specify the taste index in (\ref{qmass}). Notice the flavor singlet field 
$\Phi$ is rendered heavy by 
the $U(1)_A$ anomaly and can been integrated out in \PQCPT. However,
the propagator of the flavor-neutral field deviates from a simple pole 
form \cite{Sharpe:2001fh}. Since only the valence-valence flavor-neutral
propagators are needed for our later calculations,
for $V,V' = u,d$, the 
leading-order $\eta_{V} \eta_{V'}$ propagator in the isospin limit is given by 
\cite{s&m:2002}:
\begin{equation}
{\cal G}_{\eta_{V} \eta_{V'}} =
        \frac{i \delta_{VV'}}{q^2 - m^2_{VV} +i\epsilon}
        - \frac{i}{2} \frac{\left(q^2 - m^2_{SS,I}\right)}
            {\left(q^2 - m^2_{VV} +i\epsilon \right)^{2}}\, ,
\end{equation}
One can further show that these propagators can be conveniently rewritten in a 
compact form which will be useful in our later calculations:
\begin{equation}
{\cal G}_{\eta_{V} \eta_{V'}} =
         \d_{VV'} P_V +
         {\cal H}_{VV}\left(P_V,P_V\right)\, ,
\end{equation}
where
\begin{eqnarray}
\qquad
P_V\ =\  { i \over q^2- m_{VV}^2 + i \epsilon}\, ,\ \ \ 
\cH_{VV}(A, A) 
= 
\frac{1}{2} 
\Big[\, 
(m_{SS,I}^2 - m_{VV}^2)\frac{\partial}{\partial m_{VV}^2}A \,-\,A \Big]
\label{eq:HPsdef}
\end{eqnarray}
In writing (\ref{eq:HPsdef}), we have used $A = A(m^2_{VV})$.
 
\subsection{Baryon}
As has been shown in \cite{bct0508}, to $O(\epsilon^2)$, the free Lagrangian 
for the $\bf{572}$-dimensional super-multiplet $\cB^{ijk}$ and the 
$\bf{340}$-dimensional super-multiplet $\cT_\mu^{ijk}$ fields in the 
mixed-action $SU(10|2)$ partially quenched $\chi$PT has the 
same form as in quenched and partially quenched theories 
\cite{Labrenz:1996jy,MSavage:2002,s&m:2002} with the addition of new lattice-spacing 
dependent terms:
\begin{eqnarray} \label{eqn:L}
  {\mathcal L}
  &=&
   i\left(\ol\cB v\cdot{\mathcal D}\cB\right)
  +2\a^{(PQ)}_{M}\left(\ol\cB \cB{\mathcal M}_+\right)
  +2\b^{(PQ)}_{M}\left(\ol\cB {\mathcal M}_+\cB\right)
  +2\sigma^{(PQ)}_{M}\left(\ol\cB\cB\right)\str\left({\mathcal M}_+\right)
  + a^2 \mathcal{V}_{\cB} \nonumber \\
  &-&i\left(\ol\cT^{\mu} v\cdot{\mathcal D}\cT_\mu\right)
    +\D\left(\ol\cT^{\mu}\cT_\mu\right)
    -2\g^{(PQ)}_{M}\left(\ol\cT^{\mu} {\mathcal M}_+\cT_\mu\right)
    -2\ol\sigma^{(PQ)}_{M}\left(\ol\cT^{\mu}\cT_\mu\right)\str\left({\mathcal M}_+\right) 
   + a^2 \mathcal{V}_{\cT}\, . \notag \\
\end{eqnarray}
The baryon potentials $\mathcal{V}_\cB$ and $\mathcal{V}_\cT$ in (\ref{eqn:L})
arise from the operators in $\cL^{(6)}$ of the Symanzik Lagrangian.
In the baryon Lagrangian, the mass operator is defined by:
\begin{equation}
{\mathcal M}_+ = \frac{1}{2}\left(\xi^\dagger m_Q \xi^\dagger + \xi m_Q \xi\right)
.\end{equation}
and the parameter $\D \sim \e \L_\chi$ is the mass splitting between the 
$\bf{572}$ and $\bf{340}$ in the chiral limit.
The parenthesis notation used in Eq.~\eqref{eqn:L} is that 
of~\cite{Labrenz:1996jy}. Further, the embeding of the octet and decuplet 
baryons in their super-multiplets is the same as before 
\cite{MSavage:2002,s&m:2002}.
The Lagrangian describing the interactions of the $\cB^{ijk}$ 
and $\cT_\mu^{ijk}$ with the pseudo-Goldstone mesons in the mixed-action is 
again the same as in the partially quenched theories \cite{s&m:2002} with a 
new $a^2$-dependence term:
\begin{eqnarray} \label{eqn:Linteract}
  {\cal L} &=&   
	  2 \a \left(\ol \cB S^{\mu} \cB A_\mu \right)
	+ 2 \b \left(\ol \cB S^{\mu} A_\mu \cB \right)
	+ 2{\mathcal H}\left(\ol{\cT}^{\nu} S^{\mu} A_\mu \cT_\nu\right) \nonumber \\ 
    	&+& \sqrt{\frac{3}{2}}\cC
  		\left[
    			\left(\ol{\cT}^{\nu} A_\nu \cB\right)+ \left(\ol \cB A_\nu \cT^{\nu}\right)
  		\right] + a^2 \cL^{(6)}_{int} \,   
\label{interL}
.\end{eqnarray}
The axial-vector and vector meson fields $A_\mu$ and $V_\mu$
are defined by: $ A^{\mu}=\frac{i}{2}
\left(\xi\partial^{\mu}\xi^\dagger-\xi^\dagger\partial^{\mu}\xi\right)$  
and $V^{\mu}=\frac{1}{2} \left(\xi\partial^{\mu}\xi^\dagger+\xi^\dagger\partial^{\mu}\xi\right)$.
The latter appears in  Eq.~\eqref{eqn:L} for the
covariant derivatives of $\cB_{ijk}$ and $\cT_{ijk}$ 
that both have the form
\begin{equation}
  ({\mathcal D}^{\mu} \cB)_{ijk}
  =
  \partial^{\mu} \cB_{ijk}
  +(V^{\mu})^{l}_{i}\cB_{ljk}
  +(-)^{\eta_i(\eta_j+\eta_m)}(V^\mu)^{m}_{j}\cB_{imk}
  +(-)^{(\eta_i+\eta_j)(\eta_k+\eta_n)}(V^{\mu})^{n}_{k}\cB_{ijn}
.\end{equation}
The vector $S^{\mu}$ is the covariant spin operator \cite{Jenkins:1991jv,
Jenkins:1991es} and $a^2\cL^{(6)}_{int}$ is from the operators in 
$\cL^{(6)}$ of the Symanzik Lagrangian. As we will see later, the explicit 
form of this term is not required in our calculations. The effective 
axial-vector current from $a^2\cL^{(6)}_{int}$ can be obtained by 
a simple argument.  
The parameters that appear in the mixed-action \PQCPT\ Lagrangian can be 
related to those in \CPT\ by matching. Further, since 
QCD is contained in the fourth-root of the sea-sector of the theory, 
one should restrict oneself to one taste for each flavor of 
staggered sea quark when performing the matching. To be more specific, 
one restricts to the $q_{S}q_{S}q_{S}$ sector \cite{bct0508} and compares the 
mixed-action \PQCPT\ Lagrangian obtained with the $\chi$PT. With this identification and matching procedure, one finds \cite{s&m:2002}:
\begin{eqnarray}
\alpha_M & = & {2\over 3}\alpha_M^{(PQ)} - {1\over 3} \beta_M^{(PQ)}
\, ,\quad \
\sigma_M \ =\ \sigma_M^{(PQ)} + {1\over 6}\alpha_M^{(PQ)}
+ {2\over 3}\beta_M^{(PQ)}
\nonumber\\
\gamma_M & = & \gamma_M^{(PQ)}
\, , \quad \
\overline{\sigma}_M\ =\ \overline{\sigma}_M^{(PQ)}
\, .
\label{eq:mequals}
\end{eqnarray}
 Further, when restricting to the tree level, from (\ref{eqn:Linteract}) one 
also finds \cite{s&m:2002}:
\begin{eqnarray}
\alpha & = & {4\over 3} g_A\ +\ {1\over 3} g_1
\, ,\ \ \ 
\beta \ =\ {2\over 3} g_1 - {1\over 3} g_A
\, ,\ \ \ 
{\cal H} \ =\ g_{\Delta\Delta}
\, ,\ \ \ 
{\cal C} \ =\ -g_{\Delta N}
\, ,
\label{eq:axrels}
\end{eqnarray}   
and $g_X = 0$.   

\section{The Axial-Vector Current} 
The matrix element of the axial-vector current, 
$\overline{q}\tau^a\gamma_\mu\gamma_5 q$, have been studied extensively both 
on the lattice \cite{Edward:2005,Blum:2004,Khan:2005} and 
$\chi$PT \cite{Jenkins:1991jv,Jenkins:1991es,Borasoy:1999,Zhu:2001,
s&m:2002,s&m:2004}. 
For our convention, we will follow \cite{bct0412,bct0504} and use the 
following charge matrix for the flavor-changing current in extending the 
isovector axial current $\overline{Q}\overline{\tau}^a\gamma_\mu\gamma_5 Q$ 
to PQQCD since we are interested in the neutron to proton axial 
transition:
\begin{equation}
\overline{\tau}^{+} = (0,1,0,0,0,....,0)\, . 
\end{equation}
With this convention, at leading order, the flavor-changing axial current is 
given as \cite{s&m:2002}:
\begin{eqnarray}
^{(PQ)}j_{\mu,5}^{+}
& &\rightarrow\, 
2\alpha\ \left(\overline{\cal B} S_\mu {\cal B}\ {\overline{\tau}^{+}_{\xi +}}\right)
\ +\ 
2\beta\ \left(\overline{\cal B} S_\mu\ {\overline{\tau}^{+}_{\xi +}}{\cal B} \right)
\ +\  
2{\cal H} \left(\overline{\cal T}^\nu S_\mu\ {\overline{\tau}^{+}_{\xi +}}{\cal T}_\nu \right)
\nonumber\\
& &  
\ +\ 
\sqrt{3\over 2}{\cal C} 
\left[\ 
\left( \overline{\cal T}_\mu\ {\overline{\tau}^{+}_{\xi +}} {\cal B}\right)\ +\ 
\left(\overline{\cal B}\ {\overline{\tau}^{+}_{\xi +}} {\cal T}_\mu\right)\ \right]
\ \ + \ a^2 j^{+}_{a \mu,5}\, \ldots.
\label{eq:LOaxialcurrent}
\end{eqnarray}
where $ \overline{\tau}^{+}_{\xi +}\ =\ {1\over 2}\left(
\xi\overline{\tau}^{+}\xi^\dagger
+\xi^\dagger\overline{\tau}^{+}\xi\right)$ and $j^{+}_{a \mu,5}$ is obtained 
from $\cL^{(6)}_{int}$ in (\ref{eqn:Linteract}). Notice since the insertion of 
the mass matrix will be at $O(\epsilon^4)$, we only need to take the tree 
level contribution from $a^2 J^{+}_{a \mu, 5}$ into our calculation. 
Therefore, although there are a lot of terms in the axial current arise
from the $a^2$ corrections to the lattice operator itself,  
the net contributions to the current 
$j^{+}_{a \mu,5}$ can be effectively written as:
\begin{equation}
j^{+}_{a \mu,5}\,  \overset{\text{eff}}{=} \, 2C_{a}\ \left(\overline{\cal B} S_\mu {\cal B}\ {\overline{\tau}^{+}_{\xi +}}\right)
\ +\ 
2C'_{a}\ \left(\overline{\cal B} S_\mu\ {\overline{\tau}^{+}_{\xi +}}{\cal B} \right)\, .
\end{equation}

\begin{figure}
\includegraphics[width=0.5\textwidth]{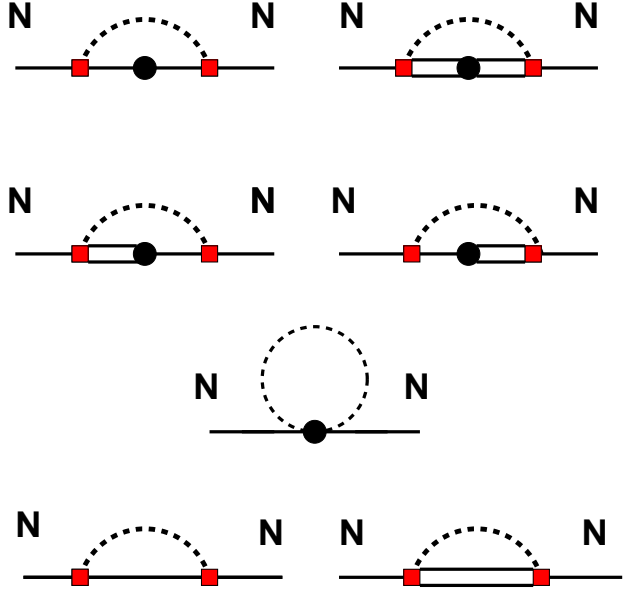}
\caption{The one-loop diagrams which contribute to the leading non-analytic 
terms of the nucleon axial transition matrix element. Mesons are represented
by a dashed line while the single and double lines are the symbols 
for a nucleon and a decuplet respectively. 
The solid circle is an insertion of the axial current operator and the
solid squares are the couplings given in (\ref{interL}). 
The wave function renormalization diagrams are depicted by the two
diagrams at the bottom row.}
\label{fig0}
\end{figure}

The calculation of the matrix element of the 
neutron to proton axial transition at next-to-leading order is 
$O(\epsilon^3)$ in our power counting. These leading non-analytic terms 
are from the one-loop diagrams shown in Fig \ref{fig0}. To obtain the 
complete $O(\epsilon^3)$ calculation, from our power counting scheme:
$\epsilon^2 \sim a^2\Lambda^2_QCD$ and 
$\epsilon^2 \sim m_{q}/\Lambda_{QCD}$, we see that in addition to taking 
the $a^2$-dependence of the loop meson masses into our calculation,  
we must evaluate the $j^{+}_{a \mu,5}$ at tree level. After carrying 
out the calculation, one finds:
\begin{eqnarray}
\langle p |j^{+}_{\mu,5} | n\rangle
& = & \Big [\ 
\Gamma_{pn} \,+\, c_{pn} \ \Big ] \ 2 \overline{U}_p S_\mu U_n\,,
\label{eq:axmat}
\end{eqnarray}
where $c_{pn}$'s are from the contributions of 
local counterterms involving one insertion of mass matrix 
$m_Q$\footnote{These local terms have the expressions 
$a_{1}m^2_{VV} + a_{2}m^2_{SS}$ with $a_{1}$ and $a_{2}$ must being
determined from the lattice calculations} 
and $\Gamma_{pn}$ is given by:
\begin{eqnarray}
\qquad \qquad \Gamma_{pn} & = &\,  g_A + a^2C' - \frac{4}{3f^2}\,\Bigg [ \,\, g_A^3\frac{3}{2}\Big 
(\,2R_{1}(m_{VS},\mu)\,+\,2A(m_{SS,I})\, \Big )\,\,\qquad \qquad 
\nonumber \\
&-& g_1^3\frac{1}{8}\Big (\, R_{1} (m_{VV},\mu)\,-\,R_{1}(m_{VS},\mu)\, \Big)\,+\,\frac{3}{2}g_A R_{1}(m_{VS},\mu) \, \nonumber \\
&-& g_A^2g_{1}\Big 
(\,2R_{1} (m_{VV},\mu)\,-\,2R_{1}(m_{VS},\mu)\,-\,6A(m_{SS,I}) \Big )\,
\nonumber \\
&-& g_Ag_1^2\Big 
(\,\frac{17}{8} R_{1} (m_{VV},\mu)\,-\,\frac{17}{8}R_{1}(m_{VS},\mu)\,-\,3A(m_{SS,I}) 
\Big )
\nonumber\\
& - & g_{\Delta N}^2g_{1}\frac{4}{9}\Big 
(\, N_{1} (m_{VV},\Delta, \mu)\,-\, N_{1}(m_{VS},\Delta,\mu)\, \Big )\,
\nonumber \\
& + & g_{\Delta N}^2g_{\Delta \Delta}\frac{60}{81}\Big 
(\,J_{1} (m_{VV},\Delta,\mu)\,+\,\frac{2}{3} J_{1}(m_{VS},\Delta,\mu)\, 
\Big ) \nonumber\\
& - & g_{\Delta N}^2g_{A}\Big 
(\, \frac{16}{9}\Big \{ N_{1} (m_{VV},\Delta, \mu) +  N_{1} (m_{VS},\Delta, \mu)\Big \} \nonumber \\
&-&2 \Big \{J_{1}(m_{VV},\Delta,\mu) + J_{1}(m_{VS},\Delta,\mu) \Big \}\, \Big )\, \Bigg ]
\label{eq:npaxial1}
\end{eqnarray}
where $C' = C_{a} + C'_{a}$, the functions $J_1(m,\Delta,\mu)$, $R_1(m,\mu)$, $N_1(m,\Delta,\mu)$
are defined in the Appendix, $\Delta$ is the $\Delta$-nucleon mass 
splitting, $m_{ab}$ are given in (\ref{qmass}) and lastly, the function 
$A(m_{SS,I})$ is given by:
\begin{equation}
A(m_{SS,I}) = \frac{1}{32\pi^2}\Big(m^2_{SS,I}-m^2_{VV}\Big)\Big[\,\log \frac{m^2_{VV}}{\mu^2}\,+\,1\,\Big ]
\label{eq:A(m,a)}
\end{equation} 

\begin{figure}
\includegraphics[width=0.6\textwidth]{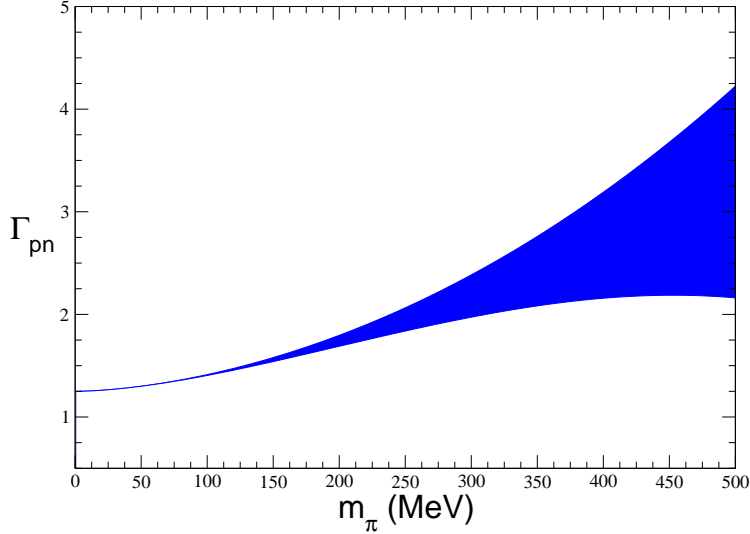}
\caption{$\Gamma_{pn} \rightarrow g_A$ in the chiral limit. Here the physical
value of $g_A$ is set to be $1.25$. The band structure indicates the
variation of $g_A$ due to the bounds on $g_{\Delta \Delta}$ and $g_{\Delta N}$.}
\label{fig1}
\end{figure}

All of the couplings in eq.~(\ref{eq:npaxial1}) 
take their chiral-limit values. It is easy to see that all the 
$a^2$-dependence in (\ref{eq:npaxial1}) is contained in $m_{VS}$,  
$m_{SS,I}$ and $a^2C'$. Notice with our $R_{1}(m,\mu)$,
$J_{1}(m,\Delta,\mu)$ and 
$N_{1}(m,\Delta,\mu)$ we have subtracted off the chiral and continuum limit
values of the loop diagrams by hand. This corresponds to a renormalization
of the tree level coefficients, and produces $g_A$ which is the chiral
limit value (Fig \ref{fig1}). 

\begin{figure}
\includegraphics[width=0.6\textwidth]{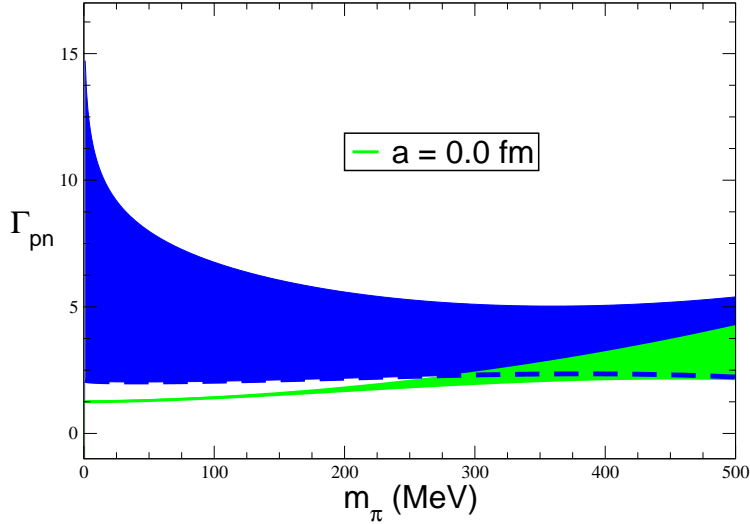}
\caption{In this figure, we have plotted $\Gamma_{pn}$ as a function of 
$m_{\pi}$. The green band stands for the band with $a = 0$
while the blue band are the $\Gamma_{pn}$ 
obtained from $g_A = 1.25$, $C_3 + C_4 = 2.34{\text{fm}}^{-6}$,
reasonably known bounds on $g_1$, 
$g_{\Delta N}$ and $g_{\Delta \Delta}$:
$-1 \leq g_{1} \leq 1$, $1.0 \leq |g_{\Delta N}| \leq 2.0$, 
$2.5 \leq |g_{\Delta \Delta}| \leq 3.5$, a natural estimated value
for $C_{\text{mix}}$: 
$4.5{\text{fm}}^{-6} \leq C_{\text{mix}} \leq 9.5{\text{fm}}^{-6}$,
and finally a natural variation
of $C'$: $-3{\text{fm}}^{-2}\leq C' \leq 3{\text{fm}}^{-2}$.
The lattice spacing $a$ for the 
blue band is fixed to be $0.12{\text{fm}}$.
 }
\label{fig2}
\end{figure}

The lattice spacing dependence in Eq.~(\ref{eq:npaxial1}) 
is completely determined by three parameters, namely, 
$C_{\text{mix}}$ and $C_3+C_4$, and a new unknown low energy 
constant $C'$. In order to 
investigate the $a^2$-dependence of $np$ axial transition matrix 
element, in Fig \ref{fig2}, we have taken the unquenched limit
in which $m_{VV} = m_{SS}$\footnote
{The unphysical effects of the mixed-action PQ theory
arise from two sources. One is from the mass difference between 
valence and sea quarks and the other is because of the non-vanishing 
lattice spacing. By taking the unquenched limit, namely, 
$m^2_{VV} = m^2_{SS}$, one can eliminate part of
the unphysical effects. Notice when the lattice spacing is zero,
the physics is recovered from the PQ theory in the unquenched 
limit.} and plotted  
(\ref{eq:npaxial1}) at two different lattice spacing $a = 0.12{\text{fm}}$ 
and $a = 0$ with the pion mass varying from 
$1{\text{MeV}}$ to $500{\text{MeV}}$\footnote{We have dropped the local 
terms $c_{pn}$'s since the loop contributions
formally dominate over these terms}. 
In the figure, $g_A$ is set to be $1.25$ and the low-energy constants 
$g_{1}$, $g_{\Delta N}$, $g_{\Delta \Delta}$
are allowed to vary within their reasonably known bounds:: 
$-1 \leq g_{1} \leq 1$, $1.0 \leq |g_{\Delta N}| \leq 2.0$ and 
$2.5 \leq |g_{\Delta \Delta}| \leq 3.5$. Further, $C_{\text{mix}}$
is estimated to be 
$4.5{\text{fm}}^{-6}\leq C_{\text{mix}} \leq 9.5{\text{fm}}^{-6}$ 
\cite{bct:0607} and $C_3 + C_4$ is determined as
$C_3+C_4 = 2.34{\text{fm}}^{-6}$ \cite{Aubin:2004fs}. 
Finally, since $C'$ is assumed to be of natural size which is 
$\Lambda^2_{QCD} \sim 2.3 {\text{fm}}^{-2}$,
we have used $-3 {\text{fm}}^{-2} \leq C' \leq 3 {\text{fm}}^{-2}$ 
for the figure. 
However, keep in mind that the actual value of $C'$ must be determined from
lattice calculations.
 
From the figure, we indeed see a large lattice spacing 
dependence for $\Gamma_{pn}$. 
One might be concerned that the correction at the chiral limit is 
sufficiently large 
that the $\chi$PT prediction might be breaking down. Further,
the band at the chiral limit does not cover the value of $1.25$
which is the expected $g_A$ at $m_{\pi} \sim 0$. 
We point out that with $a = 0.12{\text{fm}}$, the 
mass square corrections to the $VS$ and $SS,I$ mesons are 
around $(400{\text{MeV}})^2$ to $(450{\text{MeV}})^2$
which implies that $\Gamma_{pn}$ at $m_{\pi} \sim 0$ on the figure
is effectively similar to that obtained at 
$m_{\pi} \sim 450{\text{MeV}}$ with $a = 0$. Therefore a large correction
is no suprise. By reducing the lattice spacing $a$ to $0.08{\text{fm}}$,
which might be the standard lattice spacing in future simulations,
the correction is around $(270{\text{MeV}})^2$.
Notice by shifting the $a = 0$ result to the left by
$450{\text{MeV}}$ units, now the green band indeed overlaps with the blue 
band at $m_{\pi} \sim 0$. 

\begin{figure}
\includegraphics[width=0.6\textwidth]{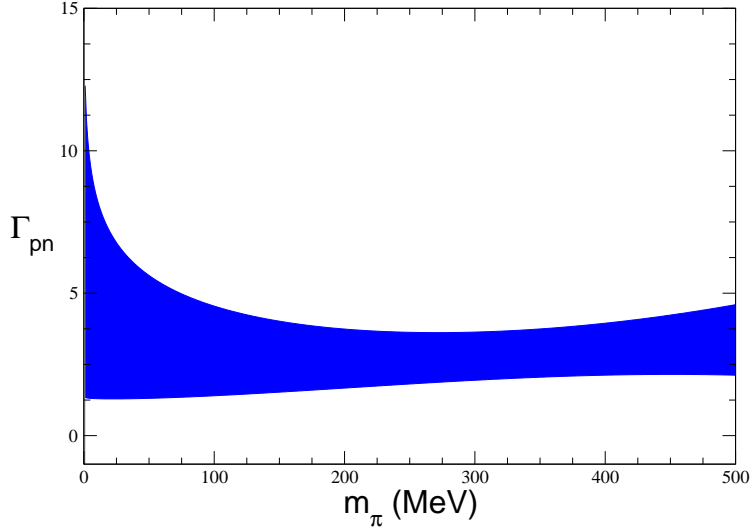}
\caption{$\Gamma_{pn}$ obtained by letting 
$C_{\text{mix}} = 0$ in (\ref{eq:npaxial1}).}
\label{fig1.6}
\end{figure}

\begin{figure}
\includegraphics[width=0.6\textwidth]{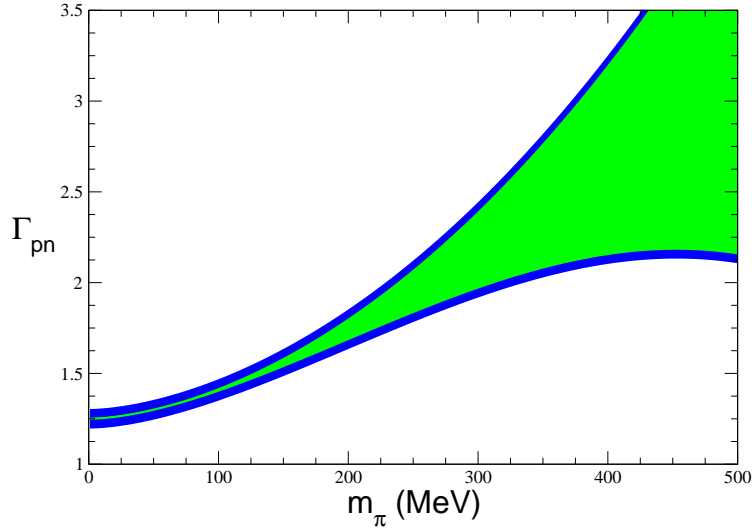}
\caption{The blue band is the $\Gamma_{pn}$ obtained by letting 
$C_{\text{mix}} = C_3 + C_4 = 0$. The green band 
represents the $\Gamma_{pn}$ with $a = 0$.}
\label{fig1.5}
\end{figure}

In addition to the large $a^2$ mass shifts,
the lack of unitarity in the PQ theory will also lead to divergences
when approaching the chiral limit and hence contributes 
a large correction to $\Gamma_{pn}$ at $m_{\pi} \sim 0$.
It is clear that the $A(m_{SS,I})$s' in Eq.~(\ref{eq:npaxial1}) are responsible
for this divergence since they are from the double pole of the
flavor neutral propagator which violate the unitarity. Indeed
if we plot Eq.~(\ref{eq:npaxial1}) as a function of $m_{\pi}$ 
with $C_{\text{mix}} = 0$, we see a large correction to $\Gamma_{pn}$ at
$m_{\pi} \sim 0$ (Fig \ref{fig1.6}). A closer look at
Eq.~(\ref{eq:npaxial1}) and Eq.~(\ref{eq:A(m,a)}) shows that the 
divergence arises from the difference of valence and taste-singlet
pion masses. In principle, $m_{VV}$ can be tuned to $m_{SS,I}$
\cite{Golterman:2005} such that the divergence is less severe. However
the large $m_{SS,I}$ will still cast doubt on the chiral expansion due
to the large mass. Also notice if
$C_{\text{mix}} = C_3 + C_4 = 0$ in Eq.~(\ref{eq:npaxial1}),
then a narrow band centered around the result obtained by setting 
$a = 0$ should be observed. Indeed this is comfirmed in Fig \ref{fig1.5}. 

Lastly, to investigate quantitatively the lattice spacing effects on $g_A$, 
let us focus on $m_{\pi} \sim 320{\text{MeV}}$ which is the 
relevant smallest pion mass used in most recent mixed-action simulations.
In Fig \ref{fig2}, we see with $a = 0.12{\text{fm}}$, $\Gamma_{pn}$ receives a 
$62$ percent correction at $m_{\pi} \sim 320{\text{MeV}}$.
This observation shows that the lattice spacing 
effects on $g_A$ of current related simulations should not be overlooked 
and that it is not clear whether $\chi$PT can be reliably used to account 
precisely for lattice spacing effects. However, with lattice spacing
$a = 0.08{\text{fm}}$, we observe a reasonable $28$ percent correction
to $\Gamma_{pn}$ at $m_{\pi} \sim 320 {\text{MeV}}$. 
We also point out that there
is an unphysical parameter $g_1$-dependence in (\ref{eq:npaxial1}) even in
the unquenched limit. This is a '' partially quenched''
artifact depending on the mixed-action and clearly from (\ref{eq:npaxial1})
the $g_{1}$-dependence will disappear when the lattice 
spacing $a$ is set to zero. 

\section{Conclusion}
The mixed-actions provide us with a powerful tool in lattice simulations 
because they enable one to use different type of fermions in the valence
and sea sectors. In particular, current lattice calcualtions using 
Ginsparg-Wilson valence quarks in staggered sea quarks can reach the
dynamical pion masses as low as $300{\text{MeV}}$. 

In this paper, we have calculated the 
$pn$ matrix element of the axial-vector current up to $O(\epsilon^3)$ order 
using the mixed-action of Ginsparg-Wilson 
valence quarks in staggered sea quarks. Further, we have detailed the 
lattice spacing artifacts for this matrix element. To $O(\epsilon^3)$,
we found that the $pn$ axial-current matrix element depends on the lattice
spacing via three parameters, namely, $C_{{\text{mix}}}$, $C_{3} + C_{4}$ and 
finally a new low energy constant $C'$. The low energy constant 
$C_{{\text{mix}}}$ affects the masses of mesons made from a Ginsparg-Wilson 
quark and a staggered quark. Its physical value can be fixed either from
mixed meson masses or the pion charge radius \cite{bct:0607}. Further, the
combination of parameters $C_3+C_4$ has been already constrained from 
staggered meson lattice data \cite{Aubin:2004fs}. The new low energy constant 
$C'$, on the other hand, can be evaluated from the lattice simulations 
of determining the nucleon axial charge 
\cite{Blum:2004,Khan:2005,Edward:2005}. However, as has been already 
demonstrated in \cite{bct:0607}, 
the continuum extrapolation with only one lattice spacing available will
lead to a large amount of uncertainty in the physical values of the associated
low energy constants. Ideally, this low energy
constant $C'$ will be more accurately determined if a variety of lattice 
spacings in the lattice simulations are available. 

Lastly, since the finite volume effects are small at 
the scale of dynamical quark masses and the box size available in today's 
simulations as indicated in \cite{s&m:2004,Edward:2005}, the 
formulas given here should be sufficient for the comparisons between the 
predictions from $\chi$PT and the results from numerical simulations.  

\begin{acknowledgments}
We thank Brian Tiburzi for critical discussions and reading a draft of the 
manuscript. We also thank Andreas Fuhrer for checking the figures 
and D. J. Cecile for assistance with the English in writing 
this manuscript. This work is supported in part by 
Schweizerischer Nationalfonds. 
\end{acknowledgments}

\section{Appendix}

In this Appendix, we list the functions needed in our calculations:
\begin{eqnarray}
\qquad\qquad R_{1}(m,\mu) = \frac{1}{16\pi^2}m^{2}\log\Big(\frac{m^{2}}{\mu^2}\Big)\, , \qquad\qquad\ \nonumber \\
\qquad    
\end{eqnarray}
\begin{eqnarray}
K_{1}(m,\Delta,\mu) & = & \,-\,\frac{1}{2}\frac{1}{16\pi^2}\Delta^2\log\Big(\frac{\mu^2}{4\Delta^2} \Big)\,+\,\frac{3}{4}\frac{1}{16\pi^2}\Bigg( 
\Big(m^2-{2\over 3}\Delta^2\Big)\log\Big({m^2\over\mu^2}\Big) \nonumber \\
&& \quad\,+\,  {2\over 3}\Delta \sqrt{\Delta^2-m^2}
\log\Big({\Delta-\sqrt{\Delta^2-m^2+ i \epsilon}\over
\Delta+\sqrt{\Delta^2-m^2+ i \epsilon}}\Big)
\nonumber\\
& & \quad \, +\, {2\over 3} {m^2\over\Delta} \Big(\ \pi m - 
\sqrt{\Delta^2-m^2}
\log\Big({\Delta-\sqrt{\Delta^2-m^2+ i \epsilon}\over
\Delta+\sqrt{\Delta^2-m^2+ i \epsilon}}\Big)
\Big) \Bigg )\, .
\label{eq:Kdecfun}
\end{eqnarray}
\begin{eqnarray}
J_{1}(m,\Delta,\mu) & = &\,-\,\frac{3}{2}\frac{1}{16\pi^2}\Delta^2\log\Big(\frac{\mu^2}{4\Delta^2} \Big)\,+\,\frac{3}{4}\frac{1}{16\pi^2}\Bigg(
\Big(m^2-2\Delta^2\Big)\log\Big({m^2\over\mu^2}\Big) \quad \qquad \nonumber \\
&&\quad \,+\,2\Delta\sqrt{\Delta^2-m^2}
\log\Big({\Delta-\sqrt{\Delta^2-m^2+ i \epsilon}\over
\Delta+\sqrt{\Delta^2-m^2+ i \epsilon}}\Big) \Bigg)\, .
\label{eq:decfun}
\end{eqnarray}

\end{document}